\documentclass{ws-procs975x65}%
\usepackage{amsmath}
\usepackage{amsfonts}
\usepackage{amssymb}
\usepackage{graphicx}%
\def\beq{\begin{equation}}
\def\eeq{\end{equation}}
\begin{document}

\title{From the Flamm-Einstein-Rosen bridge to the
modern renaissance of traversable wormholes}

\author{Francisco S.~N.~Lobo} 
\address{Instituto de Astrof\'isica e Ci\^encias do Espa\c{c}o, Universidade de Lisboa, \\
Faculdade de Ci\^encias, Campo Grande, PT1749-016 Lisboa, Portugal.
\\
E-mail: fslobo@fc.ul.pt
}

\begin{abstract}
We consider the possibility of multiply-connected spacetimes, ranging from the Flamm-Einstein-Rosen bridge, geons, and the modern renaissance of traversable wormholes. A fundamental property in wormhole physics is the flaring-out condition of the throat, which through the Einstein field equation entails the violation of the null energy condition. In the context of modified theories of gravity, it has also been shown that the normal matter can be imposed to satisfy the energy conditions, and it is the higher order curvature terms, interpreted as a gravitational fluid, that sustain these non-standard wormhole geometries, fundamentally different from their counterparts in general relativity. We explore interesting features of these geometries, in particular, the physical properties and characteristics of these `exotic spacetimes'.
\end{abstract}

\bodymatter

\section{Introduction}

Traversable wormholes are solutions to the Einstein field equation that violate the classical energy conditions and are primarily useful as ``gedanken-experiments'' and as a theoretician's probe of the foundations of General Relativity (GR). They are obtained by solving the Einstein field equation in the reverse direction, namely, one first considers an interesting and exotic spacetime metric, then finds the matter source responsible for the respective geometry. It is interesting to note that they allow ``effective'' superluminal travel, although the speed of light is not surpassed {\it locally}, and generate closed timelike curves, with the associated causality violations. In this rapporteur article, we consider a brief historical review ranging from the Flamm-Einstein-Rosen bridge,\cite{Flamm,Einstein-Rosen} to the geon wormhole-like structure obtained by Wheeler in 1955,\cite{geons} to the modern renaissance of wormhole physics.\cite{Morris,Visser}

\subsection{The Einstein-Rosen bridge}

Wormhole physics can originally be traced back to Flamm in 1916,\cite{Flamm} when he analyzed the then recently discovered Schwarzschild solution. One finds next that wormhole-type solutions were considered, in 1935, by Einstein and Rosen (ER),\cite{Einstein-Rosen} where they constructed an elementary particle model represented by a ``bridge'' connecting two identical sheets. This mathematical representation of physical space being connected by a wormhole-type solution was denoted an ``Einstein-Rosen bridge''.

Indeed, ER were attempting to build a geometrical model of a physical elementary ``particle'' that is finite and singularity-free. They based their discussion in terms of neutral and quasicharged ``bridges'' across a double-sheeted physical space. However, these can can easily be generalized.\cite{Visser} It is important to emphasize that at the time ER were writing, the notions of ``coordinate singularity'' and ``physical singularity'' were not cleanly separated:
It was supposed that the event horizon was the singularity. In this section we follow Ref. \refcite{Visser} closely.


The neutral Einstein-Rosen bridge is an observation that a suitable coordinate change seems to make the Schwarzschild (coordinate) singularity disappear, at $r=2M$. More specifically, ER discovered that certain coordinate systems naturally cover only two asymptotically flat regions of the maximally extended Schwarzschild spacetime. To see this, consider the ordinary Schwarzschild geometry:
\begin{equation}
ds^2=-\left( 1-\frac{2M}{r} \right)\,dt^2 + \left( 1-\frac{2M}{r} \right)^{-1}\,dr^2 + r^2 d\Omega^2 \,,
\end{equation}
followed by a coordinate change $u^2=r-2M$, so that the line element is represented by the ER form:
\begin{equation}
ds^2=-\frac{u^2}{u^2+2M}\,dt^2 + 4\left(u^2+2M \right)\,du^2 + \left(u^2+2M \right)^2\, d\Omega^2 \,,
\end{equation}
with $u\in (-\infty, + \infty)$. Note that this coordinate change discards the region containing the curvature singularity $r \in [0,2M)$. The region near $u=0$ is interpreted as a ``bridge'' connecting the asymptotically flat region near $u= + \infty$ with the asymptotically flat region near  $u=-\infty$.\cite{Einstein-Rosen}

To justify the ``bridge'' appellation, consider a spherical surface, with constant $u$, so that the area of the surface is given by $A(u) = 4\pi(2M+u^2 )^2$.
The latter possesses a minimum at $u=0$, with $A(0)= 4\pi(2M)^2$, which in modern terminology is defined as the ``throat'', while the nearby region is denoted the bridge, or the ``wormhole''. Thus, the neutral ``Einstein-Rosen'' bridge, or the ``Schwarzschild wormhole'', is identical to a part of the maximally extended Schwarzschild geometry. However, this wormhole is non-traversable, as the throat will pinch off before an observer may traverse the throat.\cite{Visser}


Relative to the quasi-charged Einstein-Rosen bridge, one starts off with the Reissner-Nordstrom metric:
\begin{equation}
ds^2=-\left( 1-\frac{2M}{r} +\frac{Q^2}{r^2} \right)\,dt^2 + \left( 1-\frac{2M}{r} +\frac{Q^2}{r^2} \right)^{-1}\,dr^2 + r^2 d\Omega^2 \,.
\end{equation}
In order to obtain the bridge construction, ER reversed the sign of the electromagnetic stress-energy tensor, which implies a negative energy-density. \cite{Einstein-Rosen} Considering $M=0$, with a coordinate change $u^2= r^2-\varepsilon^2/2$, results in:
\begin{equation}
ds^2=-\frac{u^2}{u^2+\varepsilon^2/2}\,dt^2 + 4\left(u^2++\varepsilon^2/2 \right)^2\,du^2 + \left(u^2+\varepsilon^2/2 \right)^2\, d\Omega^2 \,,
\end{equation}
which is indeed a very peculiar geometry, as it represents a massless, quasicharged object, with a negative energy density, possessing an horizon at $r= \varepsilon$ or $u=0$. In fact, this was the object that ER wished to interpret as an ``electron''.

Thus, the key ingredient of the bridge construction is the existence of an event horizon. The ER bridge is a coordinate artifact arising from choosing a coordinate patch, which is defined to double-cover the asymptotically flat region exterior to the black hole event horizon. One may easily consider generalizations of these contructions (see Ref. \refcite{Visser} for more details).

\subsection{Geons and spacetime foam}

After the pioneering work by Einstein and Rosen, in 1935, the field lay dormant for approximately two decades. In 1955, John Wheeler was beginning to be interested in topological issues in GR.\cite{geons} It is interesting to note that Wheeler considered a multiply-connected spacetime, where two widely separated regions were connected by a tunnel, which was denoted by a ``geon''. The ``geon'' concept denoted a ``gravitational-electromagnetic entity'', which are hypothesized solutions to the coupled Einstein-Maxwell field equations. In modern language the geon may be considered as a hypothetical ``unstable gravitational-electromagnetic quasisoliton''.\cite{Visser} Furthermore, Wheeler's concept can be used as the basis for building a model of nonzero charge with an everywhere ``zero charge density'', where one of the tunnel mouths will ``send forth lines of force into the space, and appear to have a charge''. While the other mouth ``must manifest an equal and opposite charge''.\cite{geons}

Essentially, two routes of research were then available. First, intensive research was dedicated to the classical dynamics of these tunnel configurations, assuming their possible existence. Second, the investigation of the quantum gravitational processes that might give rise to such configurations were explored. This led Wheeler to propose the concept of ``spacetime foam''.\cite{Visser,wheeler1}


In 1957, Misner and Wheeler presented an extensive analysis, where Riemannian geometry of manifolds of nontrivial topology was investigated with an ambitious view to explaining all of physics.\cite{Misner:1957mt} Indeed, this work was one of the first uses of abstract topology, homology, cohomology, and differential geometry in physics,\cite{Visser} and their point of view is best summarised by their phrase: ``Physics is geometry''. This is also the first paper\cite{Misner:1957mt} that introduces the term ``wormhole''. It is interesting to note that Misner and Wheeler considered that the existing well-established ``already unified classical theory'' allows one to describe in terms of empty curved space,\cite{Misner:1957mt} the following concepts: {\it gravitation without gravitation}; {\it electromagnetism without electromagnetism}; {\it charge without charge}; and {\it mass without mass} (where around the mouth of the ``wormhole'' lies a concentration of electromagnetic energy that gives mass to this region of space).

The objective of Misner and Wheeler was essentially to use the source-free Maxwell equations, coupled to Einstein gravity, in the context of nontrivial topology, to build models for classical electrical charges and all other particle-like entities in classical physics.\cite{Visser} It is now known that this geon solutions are unstable, so that classically the tunnels will collapse to form black holes, which inevitably hides the interesting multiply-connected topology behind event horizons.


Wheeler also noted the overwhelming importance of the Planck scale in gravitational physics, where at distances below the Planck length quantum fluctuations are considered to be extremely large, so that linearized theory breaks down and the quantum physics of full nonlinear Einstein gravity must be faced.\cite{Visser} Once the metric fluctuations become nonlinear and strongly interacting, one may expect that spacetime is flooded with a ``foamlike'' structure. In this context, the phrase ``spacetime foam'' is often referred to Wheeler's suggestion that the geometry and topology of space might be constantly fluctuating. An outstanding question is if these fluctuation may induce a change in topology, in order to form microscopic wormholes.

\subsection{Modern renaissance of wormhole physics}

After the geon solutions devised by Wheeler and Misner, there is a thirty year gap between their original work and the 1988 Morris-Thorne renaissance of wormhole physics.\cite{Morris} In fact, despite the fact that considerable effort was invested in attempting to understand the ``spacetime foam'' picture and the ``geon'' concept, during this period the geonlike-wormhole structures seem to have been considered a mere curiosity and were inevitably relegated to a backstage. However, isolated pieces of work did appear, such as the Homer Ellis' drainhole\cite{homerellis,homerellis2} concept and Bronnikov's tunnel-like solutions,\cite{bronikovWH} in the 1970s. It is only in 1988 that a full-fledged renaissance of wormhole physics took place, through the seminal paper by Morris and Thorne.\cite{Morris}
As effective technology, traversable wormholes are greatly lacking, and the fundamental approach was to ask the question: ``What do the laws of physics permit?'' 
%

In finding wormhole solutions, one adopts the reverse philosophy in solving the Einstein field equation, namely, one first considers an interesting and exotic
spacetime metric, then finds the matter source responsible for the
respective geometry. In this manner, it was found that some of these
solutions possess a peculiar property, namely ``exotic matter'',
involving a stress-energy tensor that violates the null energy
condition (NEC). These geometries also allow closed timelike curves, with the
respective causality violations.\cite{Visser,Morris:1988tu,Lobo:2010sz} Another interesting feature of these spacetimes is that they allow ``effective'' superluminal
travel, although, locally, the speed of light is not surpassed.
These solutions are primarily useful as ``gedanken-experiments'' and
as a theoretician's probe of the foundations of GR, and will be extensively reviewed throughout this review paper.\cite{Visser,Lobo:2007zb}

\section{Wormhole physics}

\subsection{Static geometry: Spacetime metric and field equations}

Consider the following static and spherically symmetric wormhole
solution
\begin{equation}
ds^2=-e ^{2\Phi(r)} \,dt^2+\frac{dr^2}{1- b(r)/r}+r^2 \,(d\theta
^2+\sin ^2{\theta} \, d\phi ^2) \,, \label{metricwormhole}
\end{equation}
where $\Phi(r)$ and $b(r)$ are arbitrary functions of the radial
coordinate $r$. $\Phi(r)$ is denoted the redshift function, for it
is related to the gravitational redshift, and $b(r)$ is denoted the
shape function, as can be shown by embedding diagrams, it
determines the shape of the wormhole.\cite{Morris} 
For the wormhole to be traversable it must have
no horizons, which implies that $g_{tt}=-e^{2\Phi(r)}\neq 0$, so
that $\Phi(r)$ must be finite everywhere.
The coordinate $r$ is non-monotonic in that it decreases from $+\infty$ to a
minimum value $r_0$, representing the location of the throat of the
wormhole, where $b(r_0)=r_0$, and then it increases from $r_0$ to
$+\infty$. Although the metric coefficient $g_{rr}$ becomes divergent at the throat, which is signalled by the coordinate singularity, the proper radial distance $ l(r)=\pm\,\int_{r_0}^r{[1-b(r)/r]^{-1/2}}\,dr$ is required to be finite everywhere. The proper distance decreases from $l=+\infty$, in the upper universe, to $l=0$ at the throat, and then from zero to $-\infty$ in the lower universe.


Using the Einstein field equation, $G_{\mu\nu}=8\pi
\,T_{\mu\nu}$ (with $c=G=1$), we obtain the following stress-energy scenario
\begin{eqnarray}
\rho(r)&=&\frac{1}{8\pi} \;\frac{b'}{r^2}   \label{rhoWH}\,,\\
p_r(r)&=&\frac{1}{8\pi} \left[2 \left(1-\frac{b}{r}
\right) \frac{\Phi'}{r} -\frac{b}{r^3}\right]  \label{prWH}\,,\\
p_t(r)&=&\frac{1}{8\pi} \left(1-\frac{b}{r}\right)\left[\Phi ''+
(\Phi')^2- \frac{b'r-b}{2r(r-b)}\Phi'
-\frac{b'r-b}{2r^2(r-b)}+\frac{\Phi'}{r} \right] \label{ptWH}\,,
\end{eqnarray}
where $\rho(r)$ is the energy density, $p_r(r)$ is the radial
pressure, and $p_t(r)$ is the lateral pressure measured in the
orthogonal direction to the radial direction. Using the
conservation of the stress-energy tensor,
$T^{\mu\nu}{}_{;\nu}=0$, we obtain the
following equation
\begin{equation}
p_r'=\frac{2}{r}\,(p_t-p_r)-(\rho +p_r)\,\Phi '
\label{prderivative} \,,
\end{equation}
which can be interpreted as the relativistic Euler equation, or
the hydrostatic equation of equilibrium for the material
threading the wormhole. Note that one now has three equations with five unknown functions of the radial coordinate. Several strategies to solve these equations are available, for instance, one can impose an equation of state,\cite{Sushkov:2005kj,Lobo:2005us,Lobo:2005yv,Lobo:2005vc,Lobo:2006ue} and consider a specific choice of the shape function or of the redshift function.

To be a solution of a wormhole, one needs to impose that the throat flares out. Using embedding diagrams,\cite{Morris,Misner} this flaring-out condition entails the following condition 
\begin{equation}
\frac{d^2r}{dz^2}=\frac{b-b'r}{2b^2}>0   \label{flareout}\,.
\end{equation}
At the throat, we verify that the shape function satisfies the
condition $b'(r_0)<1$. Note that the above treatment has the drawback of being coordinate dependent. For a covariant treatment, we refer to the analysis outlined in Refs. [\refcite{hochvisserPRL98,Hochberg1}]. We will see below that this condition plays a fundamental role in the analysis of the violation of the energy conditions. 

\subsection{The violations of the energy conditions}

\subsubsection{Classical pointwise energy conditions}

The NEC asserts that for {\it any} null vector $k^{\mu}$, we have 
$T_{\mu\nu}k^{\mu}k^{\nu}\geq 0$. For a diagonal stress-energy tensor, i.e., $T_{\nu }^{\mu }=\mathrm{diag} \left[-\rho(r),\; p_{r}( r), \; p_{t}(r),\; p_{t}(r)\right]$, this implies 
$\rho+p_{r}\geq 0$ and $\rho+p_{t}\geq 0$. Using the Einstein field equations (\ref{rhoWH}) and (\ref{prWH}), evaluated at the throat $r_0$, and taking into account the finite character of the redshift function so that $(1-b/r)\Phi'|_{r_0} \rightarrow 0$, the flaring-out condition (\ref{flareout}) imposes the condition $(\rho+p_r)|_{r_0}<0$. This violates the NEC. In fact, it implies the violation of all the pointwise energy conditions.\cite{Visser} Although classical forms of matter are believed to obey the energy conditions, it is a well-known fact that they are violated by certain quantum fields, amongst which we may refer to the Casimir effect.

Thus, the flaring-out condition (\ref{flareout}) entails the violation of the NEC, at the throat. Note that negative energy densities are not essential, but negative pressures at the throat, $p_r(r_0)=-1/(8\pi r_0^2)$, are necessary to sustain the wormhole throat. As the violation of the energy conditions is a problematic issue, it is useful to minimize this violation. Several approaches have been explored extensively in the literature, namely, rotating solutions,\cite{teo} evolving wormhole spacetimes,\cite{kar1,kar2,Arellano:2006ex} thin-shell wormholes using the cut-and-paste procedure,\cite{Poisson:1995sv,Lobo:2003xd,Eiroa:2003wp,Lobo:2004rp,Garcia:2011aa} and modified theories of gravity.\cite{Harko:2013yb,Garcia:2010xb,MontelongoGarcia:2010xd,Capozziello:2012hr}

It is interesting to note that the violations of the pointwise energy conditions led to the averaging of the energy conditions over timelike or null geodesics.\cite{Tipler} The averaged energy conditions permit localized violations of the energy conditions, as long as on average the energy conditions hold when integrated along timelike or null geodesics. Now, as the averaged energy conditions involve averaging over a line integral, with dimensions (mass)/(area), not a volume integral, they do not provide useful information regarding the ``total amount'' of energy-condition violating matter. In order to overcome this shortcoming, the ``volume integral quantifier'' was proposed.\cite{VKarDad} Thus, the amount of energy condition violations is then the extent that these integrals become negative.

\subsubsection{Quantum Inequalities}

A new set of energy constraints was introduced by Ford and Roman in 1995, \cite{Ford1} denoted by the Quantum Inequalities (QI). Contrary to the averaged energy conditions, one does not average over the entire wordline of the observer, but weights the integral with a sampling function of characteristic width. Essentially, the inequality limits the magnitude of the negative energy violations and the time for which they are allowed to exist. Ford and Roman applied the analysis to static and spherically symmetric Morris-Thorne wormholes and concluded that either the wormhole possesses a throat size which is only slightly larger than the Planck length, or there are large discrepancies in the length scales which characterize the geometry of the wormhole.\cite{Ford2} Due to these results, Ford and Roman argued that the existence of macroscopic traversable wormholes is very improbable. But, there are a series of considerations that can be applied to the QI. Firstly, the QI is only of interest if one is relying on quantum field theory to provide the exotic matter to support the wormhole throat. But there are classical systems (non-minimally coupled scalar fields) that violate the null and the weak energy conditions,\cite{barcelovisser1} while presenting plausible results when applying the QI. Secondly, even if one relies on quantum field theory to provide exotic matter, the QI does not rule out the existence of wormholes, although they do place serious constraints on the geometry.

\subsubsection{Semi-classical and nonlinear energy conditions}

Building on the Ford-Roman QIs, Martin-Moruno and Visser propsed classical and quantum versions of a ``flux energy condition'' (FEC and QFEC),\cite{Martin-Moruno:2013sfa} based on the notion of constraining the possible fluxes measured by timelike observers. The naive classical FEC were shown to be satisfied in certain situations, and even for some quantum vacuum states, while its quantum analogue (the QFEC) was satisfied under a rather wide range of conditions.
Furthermore, Martin-Moruno and Visser presented and developed several nonlinear energy conditions suitable for usage in the semiclassical regime. 
More specifically, they considered the FEC, and novel concepts such the ``trace-of-square'' (TOSEC) and ``determinant'' (DETEC) energy conditions, and showed that these nonlinear energy conditions behave much better than the classical linear energy conditions in the presence of semiclassical quantum effects.\cite{Martin-Moruno:2013wfa} Moreover, whereas the quantum extensions of these nonlinear energy conditions seem to be quite widely satisfied as one enters the quantum realm, analogous quantum extensions are generally not useful for the linear classical energy conditions.\cite{Martin-Moruno:2015ena}

\subsubsection{Buchert averaging and energy conditions}

An interesting application of the energy conditions is to the Buchert averaging.\cite{Visser:2015mur} Note that a key feature of Buchert averaging is the realization that, either with spatial averaging or averaging over a suitably defined ensemble of spacetimes, the average of the Einstein tensor is typically not equal to the Einstein tensor of the average spacetime. The discrepancy can be viewed as an ``effective'' stress-energy, one that often violates the classical energy conditions. A particularly attractive example of this phenomenon arises when one considers spatial averages in a conformal-FLRW (CFLRW) cosmology or the ensemble average over conformal deformations of a specific FLRW geometry. These CFLRW-based models are particularly tractable, and attractive for observational reasons, as the CMB is not distorted. Furthermore, it is possible to prove some rigorous theorems regarding the interplay between Buchert averaging and the classical energy conditions. We refer the reader to Ref. \refcite{Visser:2015mur} for more details.

\subsubsection{Two roads to the null energy condition}

The null energy condition has sweeping consequences in GR. However, it has been argued that it has been misunderstood as a property of matter,\cite{Parikh:2015wae} when in fact it is better viewed as a constraint on spacetime geometry. In fact, the geometric formulation of the NEC was derived from worldsheet string theory, where it is simply the Virasoro condition for a closed string moving in a curved background. Furthermore, it was shown that there is an entirely different thermodynamic origin of the NEC, if gravity emerges from some holographic theory. Thus, rather than being an incidental property of matter, it was argued in Ref. \refcite{Parikh:2015wae} that the validity of the NEC appears to hint at the origins of gravity.

\subsubsection{Probing faster than light travel and chronology protection with superluminal warp drives}

It is interesting that wormholes are not the only geometries that violate the energy conditions, and it was shown that superluminal spacetimes, such as the {\it warp drive}, also violate the weak energy condition (WEC).\cite{Alcubierre:1994tu,Lobo:2004wq} Indeed, while GR ranks undoubtedly among the best theories of physics ever developed, it is also among those with the most striking implications. In particular, GR admits solutions which allow faster than light motion, and consequently effective time travel. It was shown that a ``pre-emptive'' chronology protection mechanism destabilises superluminal warp drives via quantum matter back-reaction, and therefore forbids even the conceptual possibility to use these solutions for building a time machine.\cite{Finazzi:2009jb,Liberati:2016brg} This result was considered both in standard quantum field theory in curved spacetime as well as in the case of a quantum field theory with Lorentz invariance breakdown at high energies.

\subsection{Evolving wormholes and flashes of WEC violations}

Its interesting to note that evolving wormhole spacetimes may involve ``flashes'' of WEC violation, where the matter threading the wormhole violates the energy conditions for small intervals of time. One can consider specific cases, in which the intervals of WEC violation can be chosen to be very small.

For instance, consider the following line element of a wormhole in a cosmological
background given by
\begin{equation}\label{evolvingWHmetric}
ds^{2} = \Omega ^{2}(t) \left[- e ^{2\Phi(r)}\, dt^{2} +
{{dr^{2}}\over {1-kr^2- \frac{b(r)}{r}}} + r^2 \,\left(d\theta
^2+\sin ^2{\theta} \, d\phi ^2 \right) \right]\,,
\end{equation}
where $\Omega ^{2}(t)$ is the conformal factor, which is finite
and positive definite throughout the domain of $t$. It is also
possible to write the metric (\ref{evolvingWHmetric}) using
``physical time'' instead of ``conformal time'', by replacing $t$
by $\tau = \int \Omega (t)dt$ and therefore $\Omega (t)$ by
$R(\tau)$, where the latter is the functional form of the metric
in the $\tau$ coordinate.\cite{kar1,kar2} When the
shape function and the redshift function vanish, $b(r)\rightarrow
0$ and $\Phi(r)\rightarrow 0$, respectively, the metric
(\ref{evolvingWHmetric}) becomes the FRW metric. As
$\Omega(t)\rightarrow {\rm const}$ and $k\rightarrow 0$, it
approaches the static wormhole metric (\ref{metricwormhole}).

One can easily use the Raychaudhuri equation to verify the
non-violation of the WEC for dynamic wormholes.\cite{kar2}
Recall that the Raychaudhuri equation for a congruence of null
rays is given by
\begin{equation}
{\frac{d\hat{\theta}}{d\lambda}}=-\frac{1}{2}{\hat{\theta}}^{2}
-R_{\mu\nu}\,{k}^{\mu}\, {k}^{\nu} -2{\hat{\sigma}}^{2}
+2\hat{\omega}^2 \,,
     \label{nullRaychaud}
\end{equation}
where  $\hat{\theta}$ is the expansion of the congruence of null
rays; $\hat{\sigma}$ and $\hat{\omega}$ are the shear and
vorticity of the geodesic bundle, respectively, which are zero for
this case. From the Einstein field equation we have $R_{\mu\nu}\,
k^{\mu}\,k^{\nu}=T_{\mu\nu}\,k^{\mu}\,k^{\nu}$ for all null
$k^{\mu}$. So if $T_{\mu\nu}\,k^{\mu}\,k^{\nu} \ge 0$, we have
$\hat{\theta}^{-1}\ge {\hat{\theta}}_{0}^{-1}+\frac{\lambda}{2}$
by (\ref{nullRaychaud}). Thus if $\hat{\theta}$ is negative
anywhere it tends to $-\infty$ at a finite value of the affine
parameter, $\lambda$, i.e., the bundle must necessarily come to a
focus. For the case of a static wormhole, the expansion
$\hat{\theta}$ is given by
\begin{equation}
\hat{\theta} = \frac{2\beta}{r} \frac{dr}{dl}  \,,
\end{equation}
where $\beta$ is a positive quantity. For $l<0$, $dr/dl$ is
negative then so is $\hat{\theta}$ negative. However,
$\hat{\theta} \rightarrow -\infty$ only if $r \rightarrow 0$,
since $dr/dl$ is always finite. Therefore, either the wormhole has
a vanishing throat radius, which renders it as non-traversable, or
the WEC is violated.

For the evolving case, the expansion is given by\cite{kar2}
\begin{equation} \hat{\theta}
= \frac{2\beta}{R(\tau)}\left
(\frac{dR(\tau)}{d\tau}+\frac{1}{r}\,\frac{dr}{dl} \right)  \,.
\end{equation}
Real time $\tau$ has now been used, and as long as $dR/d\tau
>(1/r)|dr/dl|$, i.e., the wormhole is opening out fast enough, so that $\hat{\theta}$ is never negative. Thus, the fact that the bundle does not focus no longer implies that the WEC is satisfied. We refer the reader to \refcite{kar2} for specific solutions.

\subsection{Rotating wormhole solutions}

\subsubsection{The Teo solution}


Consider the stationary and axially symmetric $(3+1)-$dimensional spacetime. The latter possesses a time-like Killing vector field, which generates invariant time translations, and a spacelike Killing vector field, which generates invariant rotations with respect to the angular coordinate $\phi$. Consider the following metric\cite{teo}
\begin{equation}\label{3rwh}
  ds^2=-N^2dt^2+e^{\mu}\,dr^2+r^2K^2\left[d\theta^2
  +\sin^2\theta(d\phi-\omega\,dt)^2\right]\,,
\end{equation}
where $N$, $K$, $\omega$ and $\mu$ are functions of $r$ and
$\theta$. $\omega(r,\theta)$ may be interpreted as the
angular velocity $ d\phi/ dt$ of a particle that falls freely from
infinity to the point $(r,\theta)$. Consider the definition $e^{-\mu(r,\theta)}=1-b(r,\theta)/r$, which can be used to describe a traversable wormhole.
Assume that $K(r,\theta)$ is a positive, nondecreasing function of $r$ that determines the proper radial distance $R$, i.e., $R\equiv rK$ and $R_r>0$. We shall adopt the notation that the subscripts $_r$ and $_{\theta}$ denote the derivatives in order of $r$ and ${\theta}$, respectively.\cite{teo}

Note that an event horizon appears whenever $N=0$.\cite{teo} The
regularity of the functions $N$, $b$ and $K$ are imposed, which
implies that their $\theta$ derivatives vanish on the rotation
axis, $\theta=0,\,\pi$, to ensure a non-singular behavior of the
metric on the rotation axis. The metric (\ref{3rwh}) reduces to
the Morris-Thorne spacetime metric (\ref{metricwormhole}) in the
limit of zero rotation and spherical symmetry, $N(r,\theta)\rightarrow{\rm e}^{\Phi(r)}$, $b(r,\theta)\rightarrow b(r)$, $K(r,\theta)\rightarrow 1$ and $\omega(r,\theta)\rightarrow 0$.
%
In analogy with the Morris-Thorne case, $b(r_0)=r_0$ is identified
as the wormhole throat, and the factors $N$, $K$ and $\omega$ are
assumed to be well-behaved at the throat.
Thus, one may conclude that the metric (\ref{3rwh}) describes a
rotating wormhole geometry, with an angular velocity $\omega$. The
factor $K$ determines the proper radial distance. $N$ is the
analog of the redshift function in the Morris-Thorne wormhole and
is finite and nonzero to ensure that there are no event horizons
or curvature singularities. $b$ is the shape function which
satisfies $b\leq r$; it is independent of $\theta$ at the throat,
i.e., $b_\theta=0$; and obeys the flaring out condition $b_r<1$.

The NEC at the throat is given by
\begin{eqnarray}\label{NEC}
8\pi\,T_{\hat{\mu} \hat{\nu}}k^{\hat{\mu}} k^{\hat{\nu}}={\rm
e}^{-\mu}\mu_r{(rK)_r\over rK}
-{\omega_\theta{}^2\sin^2\theta\over2N^2}
+{(N_\theta\sin\theta)_\theta\over(rK)^2N\sin\theta}\,.
\end{eqnarray}
Rather than reproduce the analysis here, we refer the reader to
Ref. \refcite{teo}, where it was shown that the NEC is violated in
certain regions, and is satisfied in others. Thus, it is possible
for an infalling observer to move around the throat, and avoid the
exotic matter supporting the wormhole. However, it is important to
emphasize that one cannot avoid the use of exotic matter
altogether.

\subsubsection{Properties of rotating wormholes supported by phantom scalar fields and stability}

Recently, rotating wormhole solutions in GR, supported by phantom scalar fields, were further presented.\cite{Kleihaus:2014dla} It was shown that in four space--time dimensions this family of solutions evolves from the static Ellis wormhole, when a finite angular velocity is imposed at the throat. At a maximal value of the rotational velocity, the family of wormhole solutions ends in an extremal Kerr solution. The properites of these rotating wormhole solutions including their mass, angular momentum, quadrupole moment, and ergosphere were extensively discussed. In five space--time dimensions rotating wormholes with equal magnitude angular momenta were also presented. Applying stability analysis the altter, it was shown that the unstable mode of the Ellis solutions disappears, when the angular momentum of the wormhole is sufficiently high.\cite{Kleihaus:2014dla}


It is also interesting to note that for the static and spherically symmetric case, wormholes were also considered by two scalar fields.\cite{Dzhunushaliev:2015sla} These provide the possibility to obtain topologically trivial solutions in contrast with wormholes created by one scalar field. Wormholes with one scalar field are topologically non-trivial because the scalar field possesses a kink-like behaviour. The solutions with different potentials were considered, and furthermore, the profile of mass vs parameters of scalar fields was obtained.\cite{Dzhunushaliev:2015sla}

\subsubsection{Novel cosmic censorship from the Kerr-like wormhole}

The Kerr-like wormhole with phantom matter as a source, was also analyzed.\cite{Miranda:2013gqa} It was show to possess three parameters, namely, mass, angular momentum and the scalar field charge. The wormhole presented was shown to have a naked ring singularity, otherwise it is regular everywhere. The main feature of the solution is that the throat lies on a sphere of the same radius as the ring singularity, and impedes an observer to see or to reach the singularity, so that it essentially behaves as an anti-horizon. The geodesics of the wormhole were also analyzed, and it was found that an observer can traverse without consequences, however, the equator presents an infinite potential barrier which prevents the observer of reaching the throat. Furthermore, this wormhole contains a ring singularity only on the south hemisphere, without an horizon, but as in the Kerr solution, there is no manner to see the singularity. Thus, it was argued that this solution is a new kind of cosmic censorship.\cite{Miranda:2013gqa}


\subsection{Thin-shell wormholes}


Consider two distinct spacetime manifolds, ${\cal M_+}$ and ${\cal
M_-}$, with metrics given by $g_{\mu \nu}^+(x^{\mu}_+)$ and
$g_{\mu \nu}^-(x^{\mu}_-)$, in terms of independently defined
coordinate systems $x^{\mu}_+$ and $x^{\mu}_-$, respectively. The manifolds are
bounded by hypersurfaces $\Sigma_+$ and $\Sigma_-$, respectively,
with induced metrics $g_{ij}^+$ and $g_{ij}^-$.
A single manifold ${\cal M}$ is obtained by gluing together ${\cal
M_+}$ and ${\cal M_-}$ at their boundaries, i.e., ${\cal M}={\cal
M_+}\cup {\cal M_-}$, with the natural identification of the
boundaries $\Sigma=\Sigma_+=\Sigma_-$.

Now, taking into account that the interior wormhole spacetime is given by metric (\ref{metricwormhole}) and the exterior geometry is the Schwarzschild solution, the surface stresses are given by the following quantities
\begin{eqnarray}
\sigma&=&-\frac{1}{4\pi a} \left(\sqrt{1-\frac{2M}{a}+\dot{a}^2}-
\sqrt{1-\frac{b(a)}{a}+\dot{a}^2} \, \right)
    \label{surfenergy}   ,\\
{\cal P}&=&\frac{1}{8\pi a} \Bigg[\frac{1-\frac{M}{a}
+\dot{a}^2+a\ddot{a}}{\sqrt{1-\frac{2M}{a}+\dot{a}^2}}
   -\frac{(1+a\Phi') \left(1-\frac{b}{a}+\dot{a}^2
\right)+a\ddot{a}-\frac{\dot{a}^2(b-b'a)}{2(a-b)}}{\sqrt{1-\frac{b(a)}{a}+\dot{a}^2}}
\, \Bigg]         \,,
    \label{surfpressure}
\end{eqnarray}
where $\sigma$ and ${\cal P}$ are the surface energy density and
the tangential surface pressure, respectively.

The conservation equation provides us with
\begin{equation}
\sigma'=-\frac{2}{a}\,(\sigma+{\cal P})+\Xi
  \,,\label{consequation2}
\end{equation}
where $\Xi$, defined for notational convenience, is given by
\begin{eqnarray}
\Xi=-\frac{1}{4\pi a^2} \left[\frac{b'a-b}{2a\left(1-\frac{b}{a}
\right)}+a\Phi' \right] \sqrt{1-\frac{b}{a}+\dot{a}^2} \,.
       \label{H(a)}
\end{eqnarray}

The construction of dynamic shells in wormholes have been
extensively analyzed in the literature \cite{Poisson:1995sv,Lobo:2003xd,Eiroa:2003wp,Lobo:2004rp,Garcia:2011aa}, where the
stability of generic spherically symmetric thin shells to
linearized perturbations around static solutions were considered,
and applying the analysis to traversable wormhole geometries, by
considering specific choices for the shape function, the stability
regions were deduced.

\subsubsection{Negative tension branes as stable thin shell wormholes}

Negative tension branes as stable thin-shell wormholes in Reissner-Nordstroem-(anti) de Sitter spacetimes in $d$ dimensional Einstein gravity, were also investigated.\cite{Kokubu:2014vwa} Imposing $Z_2$ symmetry, traversable static thin-shell wormholes were constructed and classified in spherical, planar and hyperbolic symmetries. In spherical geometry, it was found that the higher-dimensional counterpart of Barcelo and Visser's wormholes\cite{Barcelo:2000ta} are stable against spherically symmetric perturbations. Classes of thin-shell wormholes in planar and hyperbolic symmetries with a negative cosmological constant were also found, which are stable against perturbations. In most cases, stable wormholes were found with the combination of an electric charge and a negative cosmological constant. However, as special cases, stable wormholes were found with a vanishing cosmological constant in spherical symmetry and with vanishing electric charge in hyperbolic symmetry.

\subsubsection{Thin shell collapse in CMC/maximal slicing and shape dynamics}

An interesting application of thin-shell collapse was also investigated.\cite{Gomes:2015ila} More specifically, the gravitational collapse of (massive or null) thin shells of dust in the ADM Hamiltonian formalism was studied, in a particular foliation, namely, for a constant-mean-extrinsic-curvature (CMC) in the case of a spatially compact universe, and its analogue in the case of an asymptotically flat space. Exact solutions to Einstein's equations at the nonlinear level were obtained, which take into account the backreaction of matter on geometry. The result is interesting because, in addition to providing an exact solution of GR, it also represents a solution of the newly discovered theory of Shape Dynamics.\cite{Gomes:2015ila} This theory is classically equivalent to GR, but it highlights a different (dual) symmetry to refoliation invariance, namely, spatial Weyl invariance. For this reason Shape Dynamics is expected to differ at the quantum level from the standard covariant quantization schemes of GR, and suggests that the fundamental degrees of freedom of GR are spatial conformal and diffeorphism invariant.

\section{Wormholes in modified theories of gravity}
\label{2}

Wormholes have also been extensively studied in modified theories of gravity. Without a significant loss of generality, consider the generalized gravitational field equations for a large class of modified theories of gravity, given by the following field equation\cite{Harko:2013yb}
\begin{equation}
g_1(\Psi^i)(G_{\mu\nu}+H_{\mu\nu})-g_2(\Psi^j)\,T_{\mu\nu}=\kappa^2\,T_{\mu%
\nu}\,,
  \label{generalfieldeq}
\end{equation}
where $g_i(\Psi^j)$ ($i=1,2$) are multiplicative factors that modify the geometrical sector of the field equations, and $\Psi^j$ denote generically curvature invariants or gravitational fields such as scalar fields; the term $g_2(\Psi^i)$ covers the coupling of the curvature invariants or the scalar fields with the matter stress-energy tensor, $T_{\mu\nu}$. The additional geometric term $H_{\mu\nu}$ includes the geometrical modifications inherent in the modified gravitational theory under consideration.

In order to analyse a generalized form of the energy condition, it is rather useful to rewrite this field equation as an effective Einstein field
equation, $G_{\mu\nu}=\kappa^2\, T_{\mu\nu}^{\mathrm{eff}}$, where the effective stress-energy tensor is given by
\begin{equation}
T_{\mu\nu}^{\mathrm{eff}} \equiv \frac{1+\bar{g}_2(\Psi^j)}{g_1(\Psi^i)}
\,T_{\mu\nu} -\bar{H}_{\mu\nu}\,,
\end{equation}
where $\bar{g}_2(\Psi^j)=g_2(\Psi^j)/\kappa^2$ and $\bar{H}
_{\mu\nu}=H_{\mu\nu}/\kappa^2$ are defined for notational convenience.
For this case, the violation of the generalized NEC, $T^{\mathrm{eff}}_{\mu\nu} k^\mu k^\nu < 0$, implies the following restriction
\begin{equation}
\frac{1+\bar{g}_2(\Psi^j)}{g_1(\Psi^i)}\,T_{\mu\nu} k^\mu k^\nu < \bar{H}%
_{\mu\nu}k^\mu k^\nu \,.
\end{equation}
Note that imposing $g_1(\Psi^j)=1$ , $g_2(\Psi^j)=0$, and $H_{\mu\nu}=0$, we recover GR and the standard violation of the NEC for the matter threading the wormhole, i.e., $T_{\mu\nu} k^\mu k^\nu < 0$.

If one imposes an additional condition given by $[1+\bar{g}_2(\Psi^j)]/g_1(\Psi^i)>0$, one obtains a general bound for the normal matter threading the wormhole, given by
\begin{equation}
0 \leq T_{\mu\nu} k^\mu k^\nu < \frac{g_1(\Psi^i)}{1+\bar{g}_2(\Psi^j)}\,
\bar{H}_{\mu\nu}k^\mu k^\nu \,.
\end{equation}
One may demand that the latter condition is fulfilled even if the matter stress-energy tensor satisfies the usual NEC.\cite{Harko:2013yb} One may also impose the WEC, i.e., $T_{\mu\nu}u^\mu u^\nu\geq 0$, where $u^\mu$ is the four-velocity of an observer.
In order for normal matter to satisfy the WEC, in order to have normal matter threading the wormhole, one also needs to impose the following condition
\begin{equation}
T_{\mu\nu}u^\mu u^\nu =
\frac{g_1(\Psi^i) }{\kappa^2+g_2(\Psi^j)}
\left(G_{\mu\nu} + H_{\mu\nu} \right) u^\mu u^\nu \geq 0 \,.
\end{equation}
Thus, imposing $T_{\mu\nu}u^\mu u^\nu\geq 0$ entails a restriction on the geometry arising from the modified gravity under consideration. Considering that normal matter is given by a diagonal stress-energy tensor, one can physically interpret $T_{\mu\nu}u^\mu u^\nu$ as the energy density measured by any timelike observer with four-velocity $u^\mu$. This definition is useful as using local Lorentz transformations it is possible to show that $T_{\mu\nu}u^\mu u^\nu \geq 0$ implies that the energy density is positive in all local frames of reference.

\subsection{Exact wormhole solutions with a nonminimal kinetic coupling}

In a scalar-tensor theory of gravity with a scalar field possessing the nonminimal kinetic coupling to the curvature, static and spherically symmetric solutions were considered.\cite{Sushkov:2011jh} The lagrangian of the theory contains the term $(\varepsilon g^{\mu\nu}+\eta G^{\mu\nu})\phi_{,\mu}\phi_{,\nu}$ and represents a particular case of the general Horndeski lagrangian, which leads to second-order equations of motion. The Rinaldi approach was used to construct analytical solutions describing wormholes with nonminimal kinetic coupling. It was shown that wormholes exist only if $\varepsilon=-1$ (phantom case) and $\eta>0$. Furthermore, the wormhole throat connects two anti-de Sitter spacetimes, and the metric possesses a coordinate singularity at the throat. However, since all curvature invariants are regular, there is no curvature singularity there.\cite{Sushkov:2011jh}

\subsection{Geons as wormholes of modified gravity}

As shown above, it is possible that wormholes arise as solutions of extensions of GR without violations of the energy conditions. Working in a metric-affine framework, which is experimentally supported by the physics of crystalline structures with defects,\cite{Lobo:2014nwa} explicit models supporting such solutions in four and higher dimensions were found.\cite{Olmo:2016ags,Lobo:2013prg,Lobo:2014zla,Lobo:2014nwa} It is shown that they actually represent explicit realizations of the concept of the geon introduced by Wheeler, whcih were interpreted as topologically non-trivial self-consistent bodies generated by an electromagnetic field without sources. Several of their properties were discussed, and we refer the reader to Refs. \refcite{Olmo:2016ags,Lobo:2013prg,Lobo:2014zla,Lobo:2014nwa} for more details.

\subsection{Wormholes as a cure for black hole singularities.}

Furthermore, using exactly solvable models, it was shown that black hole singularities in different electrically charged configurations can be cured.\cite{Olmo:2016hey,Bambi:2015zch} The solutions obtained describe black hole space-times with a wormhole giving structure to the otherwise point-like singularity. Furthermore, it was shown that geodesic completeness was satisfied despite the existence of curvature divergences at the wormhole throat. In some cases, physical observers can go through the wormhole and in other cases the throat lies at an infinite affine distance. The removal of singularities occurs in a non-perturbative way.\cite{Olmo:2016hey,Bambi:2015zch}

\subsection{Gravity's Rainbow and traversable wormholes}

In the context of Gravity's Rainbow, the graviton one-loop contribution to a classical energy was computed in a traversable wormhole background.\cite{Garattini:2013pha,Garattini:2015pmo} The form of the shape function considered was obtained by a linear equation of state $p=w \rho$. The approach was evaluated by means of a variational approach with Gaussian trial wave functionals. Instead of using a regularization/renormalization process, and to to handle the divergences, the distortion induced by Gravity's Rainbow was used.
The energy density of the graviton one-loop contribution, or equivalently the background spacetime, was let to evolve, and consequently the classical energy was determined. More specifically, the background metric was fixed to be Minkowskian in the equation governing the quantum fluctuations, which behaves essentially as a backreaction equation, and the quantum fluctuations were let to evolve. The classical energy, which depends on the evolved metric functions, was then evaluated. Analysing this procedure, a natural ultraviolet (UV) cutoff was obtained, which forbids the presence of an interior spacetime region, and it was argued that this may result in a multipy-connected spacetime. Thus, in the context of Gravity's Rainbow, this process may be interpreted as a change in topology, and in principle results in the presence of a Planckian wormhole.

\subsection{On wormholes creation by quantum tunnelling}

The process of quantum tunneling was studied in a self-interacting scalar field theory with non-minimal coupling to gravity.\cite{Battarra:2014naa,Battarra:2016plm} It was demonstrated that in these theories gravitational instantons can develop a neck, which is a feature prohibited in theories with minimal coupling. Furthermore, it was show that such instantons with necks lead to the materialization of bubble geometries containing a wormhole region. The relationship of neck geometries to violations of the NEC was also explored, and the bound on the size of the neck relative to that of the instanton was derived.

\subsection{Off-diagonal wormhole and black hole deformations in modified gravity theories}

General parameterizations for generic off-diagonal spacetime metrics and matter sources in GR and modified gravity were found,\cite{Vacaru:2014cwa} when the field equations decouple with respect to nonholonomic frames of reference. This allows one to construct various classes of exact solutions when the coefficients of the fundamental geometric/physical objects depend on all spacetime coordinates via corresponding classes of generating and integration functions and/or constants. Such (modified) spacetimes display Killing and non-Killing symmetries, describe nonlinear vacuum configurations and effective polarizations of cosmological and interaction constants. Certain examples of exact locally anisotropic wormholes and generic off-diagonal cosmological solutions were analysed in modified gravity, such as, in $f(R,T)$ gravity.\cite{Harko:2011kv} It was concluded that considering generic off-diagonal nonlinear parametric interactions in GR it is possible to mimic various effects in massive and/or modified gravity, or to distinguish certain classes of ``generic'' modified gravity solutions which cannot be encoded in GR.

\section{Summary and conclusion}

In this paper, we have considered a brief review of wormhole physics. It is important to emphasize that these solutions are primarily useful as ``gedanken-experiments'' and as a theoretician's probe of the foundations of general relativity. They have been extremely important in stimulating research in the issues of the energy condition violations, closed timelike curves and the associated causality violations and ``effective'' superluminal travel.
We have outlined a review dating from the ``(Flamm)-Einstein-Rosen'' bridge, the revival of the topic by Wheeler with the introduction of the ``geon'' concept in the 1960s, the full renaissance of the subject by Thorne and collaborators in the late 1980s, culminating in the monograph by Visser, and detailed the issues that branched therefrom to the present date. 

More specifically, we have presented a mathematical overview of the Morris-Thorne wormhole, paying close attention to the pointwise and averaged energy condition violations, the Quantum Inequality and modern generalizations of semi-classical nonlinear energy condition. We then, treated rotating wormholes and evolving wormholes, focussing on the energy condition violations.
Indeed, a fundamental ingredient in wormhole physics is the flaring-out condition at the throat which, in classical general relativity, entails the violation of the null energy condition. We have also presented the most general conditions in the context of modified gravity, in which the matter threading the wormhole throat satisfies all of the energy conditions, and it is the higher order curvature terms, which may be interpreted as a gravitational fluid, that support these nonstandard wormhole geometries. Thus, we explicitly show that wormhole geometries can be theoretically constructed without the presence of exotic matter, but are sustained in the context of modified gravity. Specific models were also briefly outlined.

\section*{Acknowledgments}
FSNL was supported by the Funda\c{c}\~{a}o para a Ci\^{e}ncia e Tecnologia (FCT) through the grants EXPL/FIS-AST/1608/2013, UID/FIS/04434/2013 and by a FCT Research contract, with reference IF/00859/2012.

\end{document}